\documentclass[%
 reprint,
 amsmath,amssymb,
 aps,
]{revtex4-2}

\usepackage{graphicx}% Include figure files
\usepackage{dcolumn}% Align table columns on decimal point
\usepackage{bm}% bold math
\begin{document}

\preprint{APS/123-QED}

\title{Near-field imaging of dipole emission modulated by an optical grating}% Force line breaks with \\

\author{Dong Hyuk Ko}
\author{Graham G. Brown}
\author{Chunmei Zhang}
 \email{chunmei.zhang@uottawa.ca}
\author{P. B. Corkum}
\affiliation{%
 Joint Attosecond Science Laboratory, Department of Physics, University of Ottawa, Ottawa Canada K1N 6N5\\
 National Research Council of Canada, 100 Sussex Drive, Ottawa, Canada K1A 0R6
}%

\date{\today}% It is always \today, today,

\begin{abstract}
Multiphoton-ionized electrons are born into a strong light field that will determine their short-term future. By controlling the infrared beam, we enable atoms or molecules to generate extreme ultraviolet (XUV) pulses and synthesize attosecond pulses – the shortest controlled events ever produced. Here we show that a weak obliquely incident beam imposes an optical grating on the fundamental beam, resulting in a spatially modulated attosecond pulse. We observe the modulation on a spectrally resolved near-field XUV image, encoding all information on the spectral phase of the re-collision electron and, therefore, the attosecond pulse produced by structureless atoms. Near-field imaging is an efficient method for measuring the duration of attosecond pulses, especially important for soft X-ray pulses created in Helium. For more complex systems, it allows us to measure multi-electron dynamics while creating coherent photons. In two companion papers, we show this includes autoionization and the Giant Plasmon resonance \cite{01fano, 02plasmon}. 
\end{abstract}

\maketitle

Collision physics, wherein the fate of a colliding electron is mapped from an incident beam to a scattered beam, has been central to studying multi-electron dynamics for decades \cite{03Drescher2002}. As a colliding electron interacts with its partners, it may be deflected elastically or create new particles inelastically. Attosecond science, through its precise temporal control of ionization or re-combination, can probe both possibilities while contributing the highly developed diagnostic tools of optics to collision physics. Time-resolved photoelectron spectroscopy with well-characterized attosecond pulses has been used to study Auger decay in ions or ultrafast shake-up during photoionization in atoms \cite{03Drescher2002, 04Uiberacker2007, 05RevModPhys.81.163}. The relative time delay of electrons ionized from different orbitals in atoms or different bands in solids has also been studied \cite{06Cavalieri2007}. Delay of resonant two-photon ionization has been widely investigated by applying photoelectron streaking \cite{07PhysRevLett.104.103003, 08PhysRevLett.105.263003, 09PhysRevLett.108.093001}. However, in conventional attosecond science, it is difficult to separate time delay due to multi-electron dynamics from time delay due to the electrostatic structure of a system (how the electron interacts with its charged, but static environment) since both contribute to photoelectron time delay. 

In companion papers \cite{01fano, 02plasmon}, we will show that these contributions can be separated through the sensitivity of “in situ” measurement to dynamics but its insensitivity to structure. Our purpose here is to look at in situ measurement in detail and to introduce single-image measurement. In situ methods were introduced as a simplified approach to attosecond pulse measurement in which measurement and pulse generation take place simultaneously. However, in situ methods have two weaknesses that have undermined their application. First, for chirp compensation, attosecond pulses passing through a dispersive medium after their generation cannot be measured by the in situ technique. While this should not be a significant deterrent since we are only dealing with linear dispersion, it is still disadvantageous not to measure the final beam directly. Second, in situ measurement is insensitive to the phase of the transition moment \cite{10PhysRevA.94.023825}. This is also not serious for simple systems since we measure the closely related spectral phase of the re-collision electron and we understand single photon transitions \cite{11PhysRevA.72.013401}, but, again, the final pulse is only indirectly measured.

\begin{figure*}[!ht]
	\includegraphics[width=\textwidth]{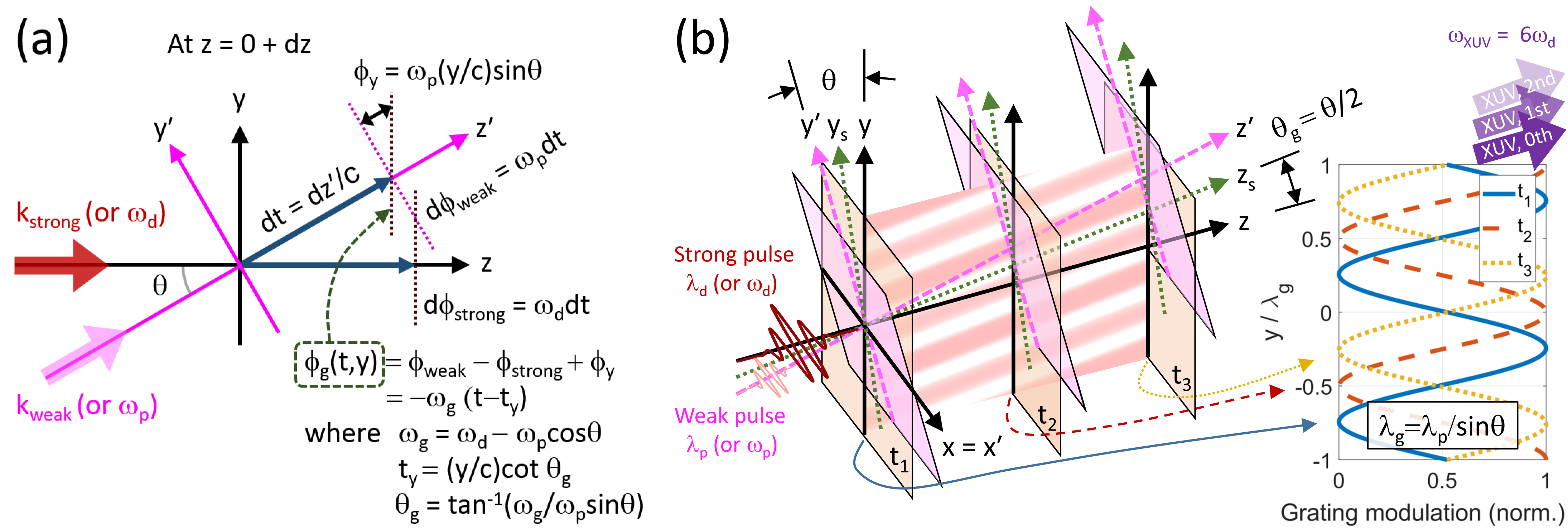}
	\caption{Schematic diagram of the obliquely overlapped two laser beams and generated XUV beams. (a) Schematic view of the laser beams and their wave fronts. Due to the angle of the weak beam, the relative phase between two beams depends on time and position, serving as a moving grating. (b) An isometric view of the co-ordinate frames for the laser and XUV beams. The electron trajectories, dominated by the intense beam, is predominantly along y and the grating is not static in this frame. The diffracted XUV beam has a frequency-shifted spectral phase $\phi(\omega_{XUV} + M\omega_g)$ where $M$ is the diffraction order. The inset shows the grating modulation along the y axis at different times. We directly image this time-dependent modulation for the near-field attosecond pulse measurement. }
\end{figure*}

However, in situ measurement gains significance when we focus on multi-electron dynamics from attosecond pulses. In any multi-electron response within the irradiated quantum system, the re-collision electron must be influenced by the interaction. This means that, for in situ measurement, any deviation from a hydrogen-like atto-chirp is a signature of multi-electron dynamics.

With the potential of multi-electron dynamic measurements in mind, here we demonstrate experimentally, supported by theory, single-image near-field measurement of an isolated attosecond pulse. The intrinsic sensitivity arises from both the all-optical nature of the measurement and from the measurement being performed near-field. With improved detection, a higher frequency driver, or a more powerful fundamental beam, our approach scales to a single-shot measurement. 

Single-image measurement of an attosecond pulse is based on the perturbative control of high harmonic generation by introducing a weak laser field in the harmonic generation medium at an angle \cite{12Dudovich2006, 13Kim2013, 14PhysRevLett.106.023001}. The perturbation, resulting in the in situ measurement, allows us to determine the electron trajectory associated with a given spectral component of an attosecond pulse. We use this to measure the attosecond pulse and strong-field dynamics.

As described mathematically in supplementary information, when two laser fields overlap non-collinearly, whether they have same frequency or not, they induce a grating that is, in general, non-stationary. A reader may worry that a grating created by same-frequency laser fields is time-independent when viewed along the bisecting angle. However, in our case one of the laser pulses is much weaker than the other. Therefore, it is appropriate to analyze the problem in the frame of the intense driving beam. In Fig. 1(a), the strong laser pulse propagates along z-axis and the weak laser pulse travels along z'-axis at an angle $\theta$ with respect to z-axis. 

When we consider an infinitesimal propagation distance $dz$ or $dz'$ for the strong and weak beams with angular frequencies of $\omega_d$ and $\omega_p$, respectively, during $dt$ at $z=0$ where the XUV is generated, the wave front of the weak beam has a phase shift with respect to that of the strong beam by $\phi_g(t, y) = -\omega_g(t-t_y)$ where $t_y = (y/c)cot\theta_g$, implying a moving grating on $z-y$ frame. Here $\omega_g = \omega_d-\omega_p cos\theta$ , meaning the angular frequency of the optical grating, $c$ is the speed of light and $\theta_g$ is a propagation angle of the optical grating defined as $\tan^{-1}(\omega_g/\omega_p \sin\theta)$. In this frame, the grating moves by time following a line of $y = z tan\theta_g$ as shown in Fig. 1(b). Consequently, the re-collision electron motion and the wave fronts of the prominent zero-order and less prominent first-order diffracted XUV beams are mainly determined by the strong laser beam but perturbed by the weak laser beam, following the momentum conservation of diffracted XUV beams as described by violet arrows in the Fig. 1(b). 

We can determine the emission times of XUV frequencies by measuring the position of the emission (that is the phase of the grating) at that frequency. The equations for the time-dependent dipole perturbed with a non-collinear beam and, numerical simulation for the spectrally resolved near-field dipole that we use to characterize an attosecond pulse, are described in supplementary information.

\begin{figure*}[t]
	\includegraphics[width=\textwidth]{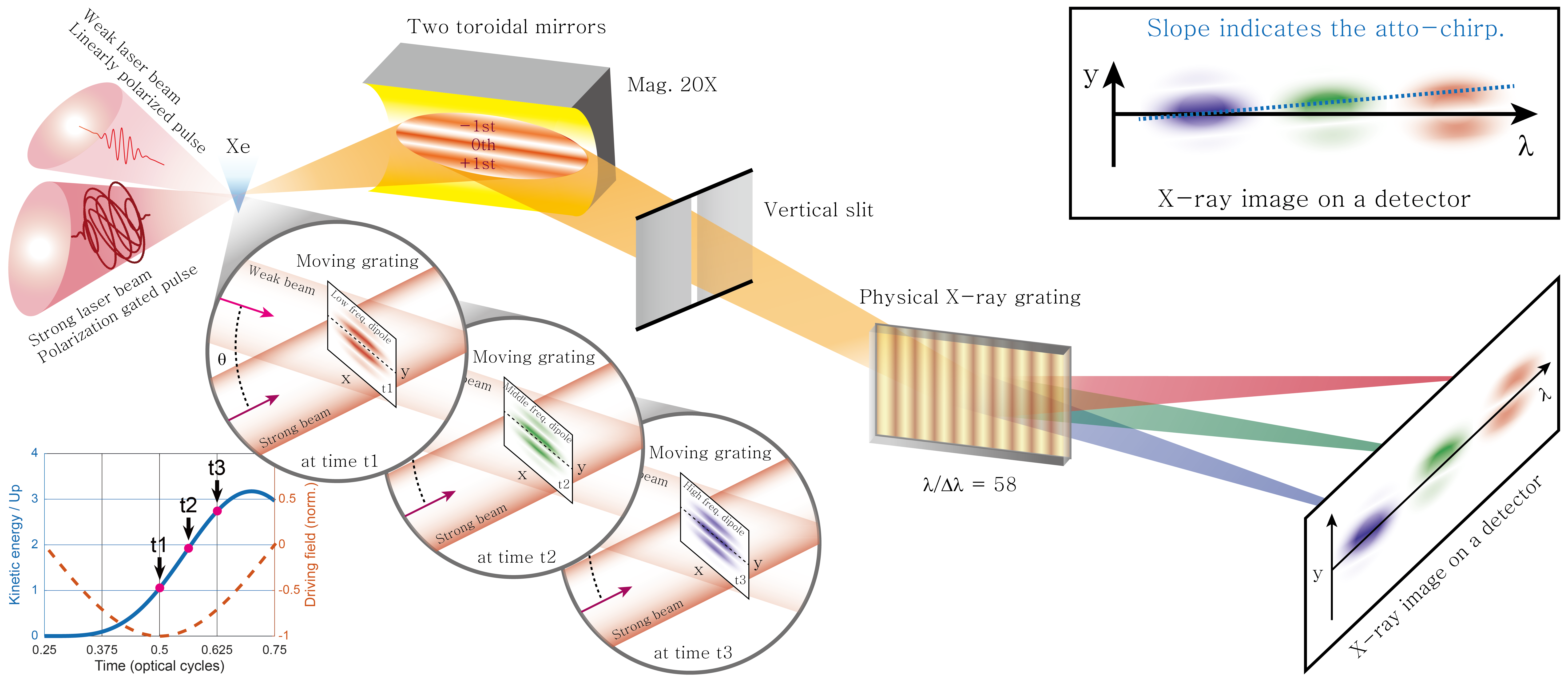}
	\caption{Schematic diagram of the obliquely overlapped two laser beams and generated XUV beams. (a) Schematic view of the laser beams and their wave fronts. Due to the angle of the weak beam, the relative phase between two beams depends on time and position, serving as a moving grating. (b) An isometric view of the co-ordinate frames for the laser and XUV beams. The electron trajectories, dominated by the intense beam, is predominantly along y and the grating is not static in this frame. The diffracted XUV beam has a frequency-shifted spectral phase $\phi(\omega_{XUV} + M\omega_g)$ where $M$ is the diffraction order. The inset shows the grating modulation along the y axis at different times. We directly image this time-dependent modulation for the near-field attosecond pulse measurement. }
\end{figure*}

To experimentally demonstrate single-image, near-field measurement of an isolated attosecond pulse, first, we generate single attosecond pulses from Xe using a carrier-envelope-phase (CEP) stabilized few-cycle laser pulse and detect the pulse by a near-field imaging XUV spectrometer, illustrated in Fig. 2 \cite{15Roy2011, 16Choi:97}. 

\begin{figure}
	\includegraphics[width=0.5\textwidth]{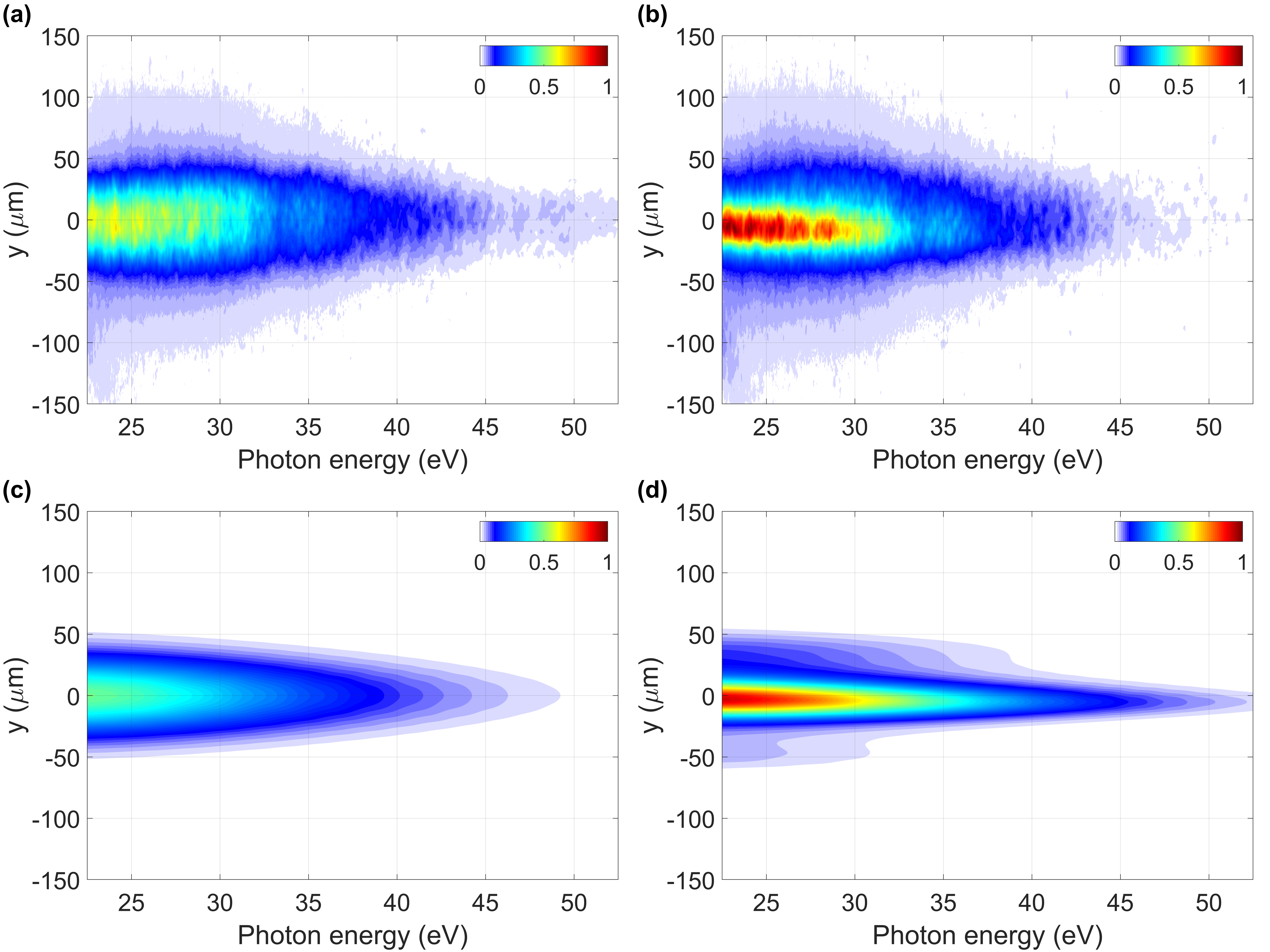}
	\caption{Measured XUV continuum spectra without/with the weak laser pulse in (a) and (b) respectively. Numerically simulated XUV continuum spectra without/with the weak laser pulse in (c) and (d) respectively by taking into account parameters in the experiment. }
\end{figure}

\begin{figure*}
	\centering
	\includegraphics[width=\textwidth]{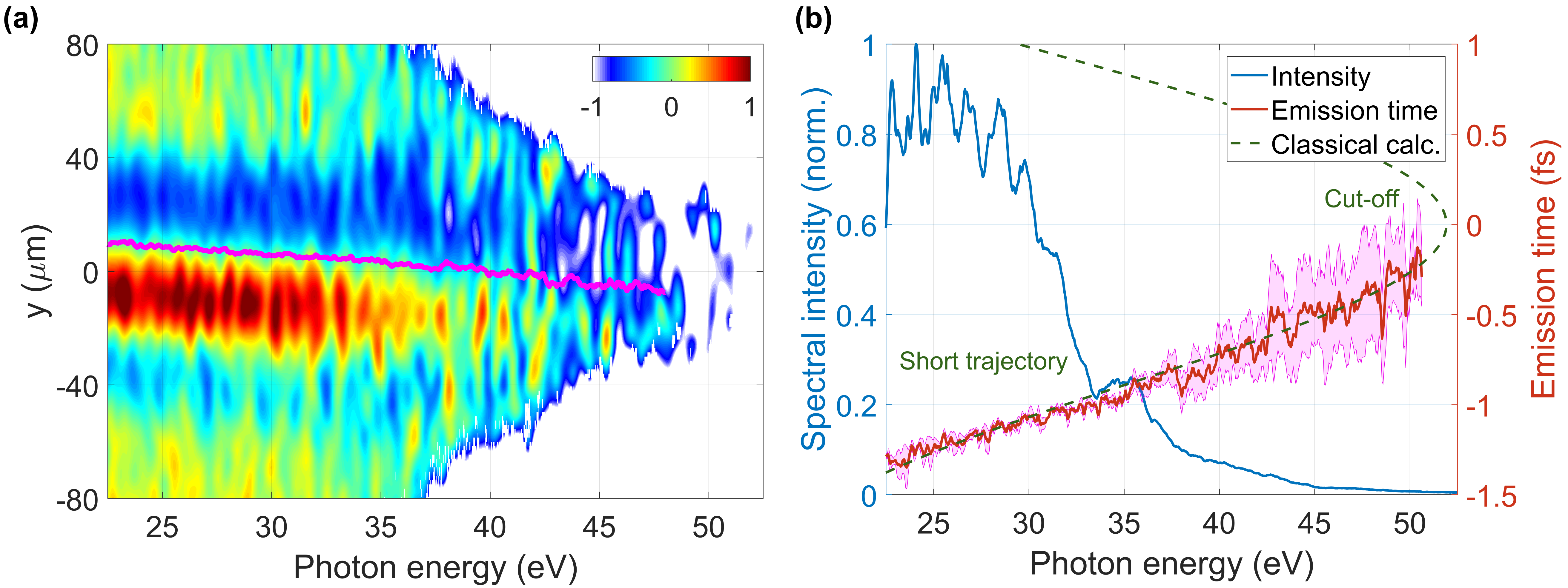}
	\caption{(a) Normalized difference of Fig. 3(b) with Fig. 3(a). The optical grating induced by the weak laser field is clearly revealed on the near-field dipole emission with 1.5-cycle modulation. The spectrally-resolved grating position is marked by a magenta line at the middle of the grating modulation for each frequency. (b) Measured spectral intensity (blue solid line) and emission time (red solid line) of the isolated attosecond pulse. The magenta shade indicates statistical error of the emission time after taking 70 measurements with 150 shots accumulated for each image. The green-dashed line is the dispersion curve of the isolated attosecond pulse calculated by a SFA model [29]. }
\end{figure*}

Figure 3(a) shows the spectrally resolved near-field dipole emission of an attosecond pulse created by applying polarization gating \cite{17Sansone443}. The peak intensity at the linearly polarized part of the pulse is $4\times10^{13}$ W cm$^{-2}$, which is strong enough to drive Xe for a broadband XUV spectrum up to 50 eV as shown in the Fig. 3(a). We confirm that the XUV source has a uniform size of 100 \textmu m from 22 eV to 30 eV at the low frequencies of the attosecond pulse, corresponding to the laser beam size. Higher energy photons near the cut-off have smaller size because the intensity to generate these frequencies is sufficiently strong over a smaller cross-section of the driving laser beam. Figure 3(c) shows the result obtained by solving the time-dependent Schrödinger equation (TDSE) with parameters chosen to match the measured near-field dipole spectrum in the Fig. 3(a). Experimental details are described in the Method section.

We then add a weak pulse at an angle of 30 mrad and intensity of $4\times10^{11}$ W cm$^{-2}$ to perturb the dipole emission, creating a grating with the period of 54 \textmu m. The grating modulation on the dipole emission is clearly pronounced in Fig. 3(b) compared with Fig. 3(a) (the same color scale). Numerical simulations using the same parameters (Fig. 3(d)) agree with the measured near-field dipole spectrum in the Fig. 3(b).

The normalized difference of Fig. 3(b) with Fig. 3(a) is plotted in Fig. 4(a). It shows that each frequency of the attosecond pulse is created with a different grating phase $\phi_g$ (thereby, a different emission time) marked by a magenta-dashed line with a scaling factor of $\lambda_g/2\pi$, implying the intrinsic atto-chirp \cite{18Salieres902, 19PhysRevLett.99.223904}. Here $\lambda_g = \lambda_p/sin\theta$ and $\lambda_p$ is the wavelength of the weak laser field. Since Xe has a smooth re-combination cross-section over the measured XUV spectrum, as confirmed by density functional theory in the supplementary information, we regard the re-combination time of the re-colliding electron as the emission time of the corresponding XUV frequency. It allows us to link re-collision and emission times (i.e. $\phi_g(\omega)=\Delta\phi(\omega)$), as explained by the equation (11) of the supplementary information. Here $\Delta \phi(\omega)=\phi(\omega+\omega_g)-\phi(\omega)$ represents the infinitesimal differentiation of the spectral phase $\phi(\omega)$ of the attosecond pulse. Due to the non-stationary optical grating, the first order has the spectral phase of $\phi(\omega+\omega_g)$, which is the spectral phase of the attosecond pulse with infinitesimal frequency shift by $\omega_g$. Accordingly, we achieve the emission time $\tau_e(\omega)$ of the attosecond pulse as $\tau_e(\omega)=\Delta\phi(\omega)/\Delta\omega=\phi_g(\omega)/\omega_g$, shown in Fig. 4(b) with a red-solid-line as well as the spectral intensity obtained from the Fig. 3(a). The shaded area in red indicates the standard deviation of the measured emission time by repeating the measurement for statistical analysis (70 images with 150 shots per image).

For the emission time in Fig. 4(b), there is only one free parameter. We set the emission time at 25.4 eV to be zero for all measurements. The emission time near cut-off shows a large deviation relative to the low energy side due to the small amplitude of the XUV spectrum. For comparison, we also plot the emission time (green-dashed-line) from the semi-classical calculation for the electron trajectories of the attosecond pulse using experimental parameters. The measured emission time agrees with the calculated value for the short trajectory, which dominates for long wavelength driver \cite{20PhysRevLett.82.1668, 21doi:10.1080/09500340.2013.765067}. 
For a real transition moment, the complete temporal characterization of the isolated attosecond pulse is accomplished by the single-image measurement, providing us the spectral amplitude and phase of the attosecond pulse. Figure 5 presents the reconstructed isolated attosecond pulse, showing a positive chirp with a duration of 320±20 as. The transform-limited-duration is 160 as. The TDSE calculation (Fig. 3(c)) also holds the full information of the attosecond pulse, overlaid by dashed lines in the Fig. 5. It agrees with the measured attosecond pulse on both the temporal intensity and phase. The shaded areas in blue and red of the Fig. 5 represent statistical errors for the temporal intensity and phase, respectively. The inset of the Fig. 5 shows the neighboring pulses at ±3 fs containing 0.2 \% of the pulse energy. 

\begin{figure}
	\centering
	\includegraphics[width=0.5\textwidth]{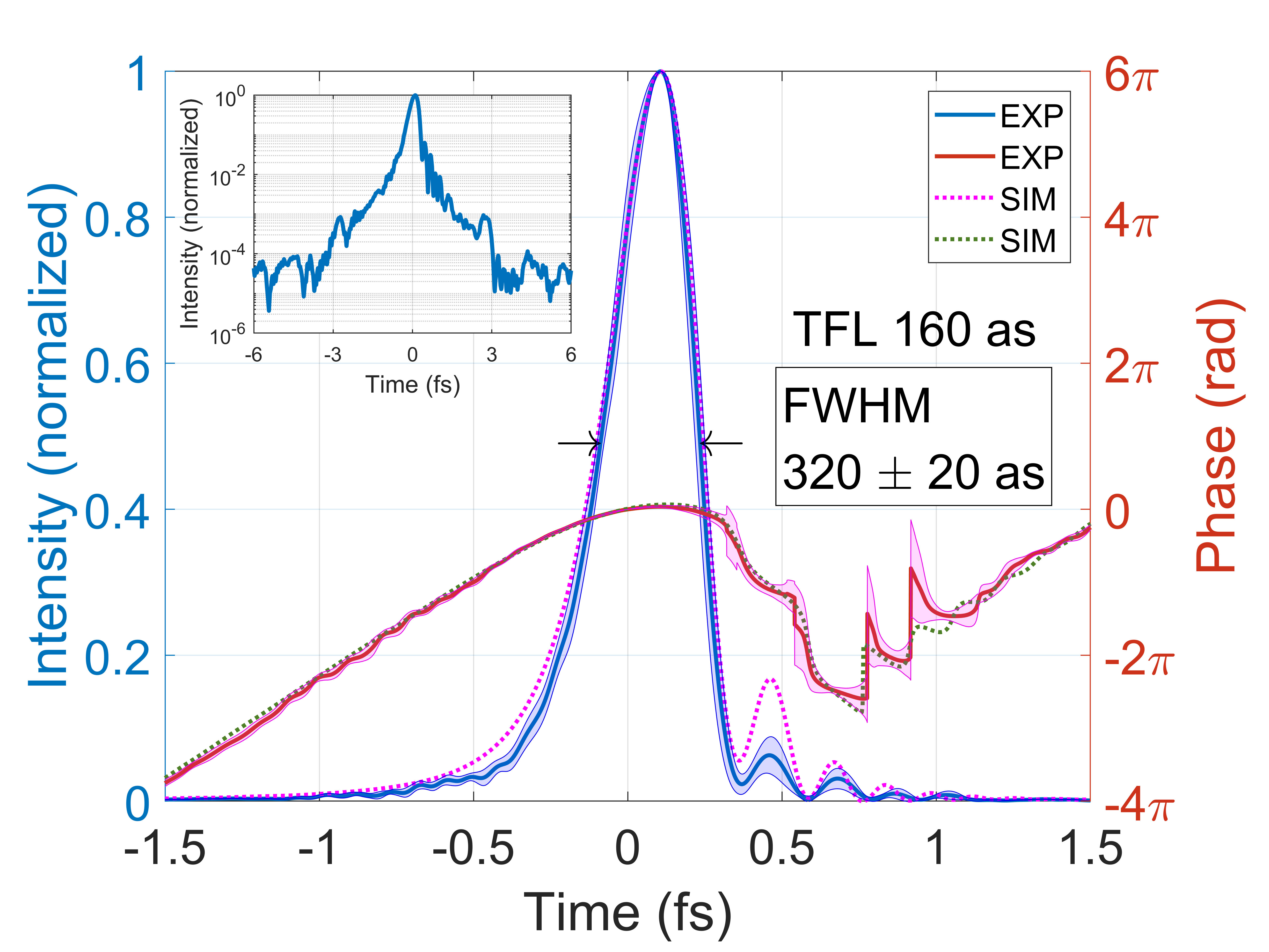}
	\caption{Complete temporal characterization of an isolated attosecond pulse obtained by taking Fourier transform of the measured spectral amplitude and phase. The shaded areas are the standard deviations of the temporal intensity (blue) and phase (red) from the statistical analysis. Dotted lines are the intensity and phase achieved by a TDSE calculation using the strong driving laser pulse only. The inset shows the measured attosecond pulse on a logarithm scale to confirm it as a single isolated attosecond pulse.}
\end{figure}

In conclusion, there are two approaches to attosecond measurement – in situ \cite{13Kim2013, 14PhysRevLett.106.023001} and ex situ \cite{22Paul1689, 23PhysRevLett.88.173903}. Here, we have extended the in situ measurement to the single image regime. We have taken our image in 300 ms (150 shots limited by our 500-Hz supersonic jet) using a 100-\textmu J 1.8-\textmu m laser operating at 1 kHz. This is the first single image technology for attosecond science. Projecting our results to 800 nm where the single atom response is stronger \cite{24PhysRevLett.103.073902}, we predict single-shot measurements using the few-cycle mJ-level lasers available in many labs. 

We have shown that in situ measurement can be performed in the near-field as well as the far-field. For near-field measurement, we have introduced an X-ray imaging system to image diffracted XUV beams from the medium where the measurement is made. The near-field approach uses photons very economically, thereby allowing attosecond pulse characterization to be extended to the water window and beyond. This follows earlier work, applying the near-field method to femtosecond autocorrelation \cite{25Hammond_2018}. However, a far-field approach, requiring measurement of the phase difference between a zero and first order of grating diffraction with respect to XUV frequency, is also possible \cite{26Bertrand2013}. 

In situ and ex situ measurements are sensitive to different (but closely related) things. In situ measurement observes the system and the dynamics in the presence of the strong laser field. Ex situ can (but does not always) use low intensity beams for measurement. For symmetric systems in the single active electron limit, in situ techniques measure the contribution of the re-colliding electron to the spectral phase \cite{10PhysRevA.94.023825}, but these methods are insensitive to the transition moment, which is readily observable with conventional streaking \cite{27Schultze1658}. Therefore, this allows in situ methods to uniquely observe multi-electron dynamics occurring during the strong-field driven re-collision process \cite{28PhysRevLett.111.233005}. In a forthcoming paper, we extend the measurement described above to show how the delay of recollision electron due to its stimulation of the plasmonic excitation becomes observable using in situ technique \cite{02plasmon}. 

\nocite{*}
\bibliography{apssamp}% Produces the bibliography via BibTeX.

\end{document}